\documentclass{aa}  

\usepackage{graphicx}
\usepackage{txfonts}
\usepackage{hyperref}
\def\reff{R_{\mathrm{e}}}

\def\fspi{f_{\mathrm{spi}}}
\def\Rspi{R_{\mathrm{spi}}}
\def\Rspilw{R_{\mathrm{spi,lw}}}

\def\Fref#1{Figure~\ref{#1}\xspace}
\def\Tref#1{Table~\ref{#1}\xspace}
\def\Eref#1{Equation~\ref{#1}\xspace}

\defcitealias{S+C21}{Paper~I}
\defcitealias{Son21a}{Paper~II}
\defcitealias{Son21b}{Paper~III}
\setcitestyle{notesep={}}

\begin{document}

   \title{The effect of spiral arms on the S\'{e}rsic photometry of galaxies}
   \titlerunning{The effect of spiral arms}
   \authorrunning{Sonnenfeld}


   \author{Alessandro Sonnenfeld\inst{1}
          }

   \institute{Leiden Observatory, Leiden University, Niels Bohrweg 2, 2333 CA Leiden, the Netherlands\\
              \email{sonnenfeld@strw.leidenuniv.nl}
             }

   \date{}

 
  \abstract
    {
The S\'{e}rsic profile is a widely used model for describing the surface brightness distribution of galaxies. 
Spiral galaxies, however, are qualitatively different from a S\'{e}rsic model.
}
   {
The goal of this study is to assess how accurately the total flux and half-light radius of a galaxy with spiral arms can be recovered when fitted with a S\'{e}rsic profile.
} 
   {
I selected a sample of bulge-dominated galaxies with spiral arms. Using photometric data from the Hyper Suprime-Cam survey, I estimated the contribution of the spiral arms to their total flux. Then I generated simulated images of galaxies with similar characteristics, fitted them with a S\'{e}rsic model, and quantified the error on the determination of the total flux and half-light radius.
}
   {
Spiral arms can introduce biases on the photometry of galaxies in a way that depends on the underlying smooth surface brightness profile, the location of the arms, and the depth of the photometric data.
A set of spiral arms accounting for $10\%$ of the flux of a bulge-dominated galaxy typically causes the total flux and the half-light radius to be overestimated by 15\% and 30\%, respectively. 
This bias, however, is much smaller if the galaxy is disk-dominated.
}
   {
Galaxies with a prominent bulge and a non-zero contribution from spiral arms are the most susceptible to biases in the total flux and half-light radius when fitted with a S\'{e}rsic profile.
If photometric measurements with high accuracy are required, then measurements over finite apertures are to be preferred over global estimates of the flux.
}

   \maketitle
%

\section{Introduction}\label{sect:intro}

\message{The column width is: \the\columnwidth}

Two widely used quantities to describe the global properties of a galaxy are the total flux and the radius enclosing half of that flux, that is, the half-light radius.
When the distance of a galaxy is known and an estimate of its stellar mass-to-light ratio is available (for example, from stellar population synthesis modelling), the total flux and the half-light radius can be converted into measurements of the total stellar mass and half-mass radius.
These are fundamental properties of a galaxy, the knowledge of which is crucial for understanding the baryon cycle in the Universe and galaxy evolution.
Stellar mass measurements can be used, for example, to constrain the efficiency of the process of star formation across cosmic time \citep{MNW18, Beh++19}, while the half-mass radius of a galaxy is sensitive to the angular momentum \citep{MMW98} and the merger history of its host dark matter halo \citep{NJO09, HNO13, Sha++14}.

Photometric and spectroscopic surveys such as the Sloan Digital Sky Survey \citep[SDSS;][]{Yor++00}, the Kilo-Degree Survey \citep[KiDS;][]{deJ++13}, the Galaxy And Mass Assembly (GAMA) survey \citep{Dri++11}, and the Hyper Suprime-Cam (HSC) Subaru Strategic Program \citep[HSC SSP, or HSC survey;][]{Aih++18} have enabled the measurement of the fluxes and sizes\footnote{Throughout this paper, I use the terms `size' and `half-light radius' interchangeably.} of millions of galaxies \citep{Sim++11,Kel++12,MVB15,Kaw++21}.
Such a large amount of data makes it possible to determine galaxy scaling relations, such as the stellar mass-size relation, with extremely high statistical precision.
When statistical uncertainties are very small, however, inferences are generally limited by systematic errors. 
This work assesses the impact of one possible source of systematic error, which is related to the method with which fluxes and sizes are determined.

The standard approach to measuring the total flux and half-light radius of a galaxy is by fitting its image with a simply parameterised model for its surface brightness distribution.
The most commonly used model is the S\'{e}rsic profile \citep{Ser68}. 
The S\'{e}rsic model consists of a smooth surface brightness distribution with three degrees of freedom in the radial direction.
The success of this model owes to the fact that it is able to approximate, on average, the surface brightness distribution of a broad range of galaxies, from massive ellipticals to disks. 
Most galaxies, however, exhibit substructure that deviates from a smooth description of the surface brightness distribution, such as spiral arms, rings, bars, or shells.
For these galaxies, a S\'{e}rsic model is unable to reproduce their images down to the noise level.
These bad fits are prone to biases. 

In this work, I investigate the possible bias introduced by fitting a S\'{e}rsic model to galaxies with spiral arms.
I focus on a particular class of objects, that of bulge-dominated spirals, also known as red spirals.
The reason for this choice is that these are objects for which it is easier to separate the spiral arms from the underlying smooth component of the surface brightness profile.
In the first part of this study, I select a sample of bulge-dominated spiral galaxies and quantify the fraction of their optical flux that is due to spiral arms.
Subsequently, I simulate images of spiral galaxies with similar characteristics to the observed sample, fit them with a S\'{e}rsic model, and measure the error on the inferred flux and half-light radius.

The structure of this work is the following.
In Sect. \ref{sect:spifrac} I show estimates of the fraction of flux in spiral arms for a select sample of galaxies.
In Sect. \ref{sect:results} I show the results of the S\'{e}rsic profile fit to the simulated galaxies.
I discuss the results and draw conclusions in Sect. \ref{sect:discuss}.


\section{Spiral arm contribution}\label{sect:spifrac}

\begin{figure*}
\includegraphics[width=\textwidth]{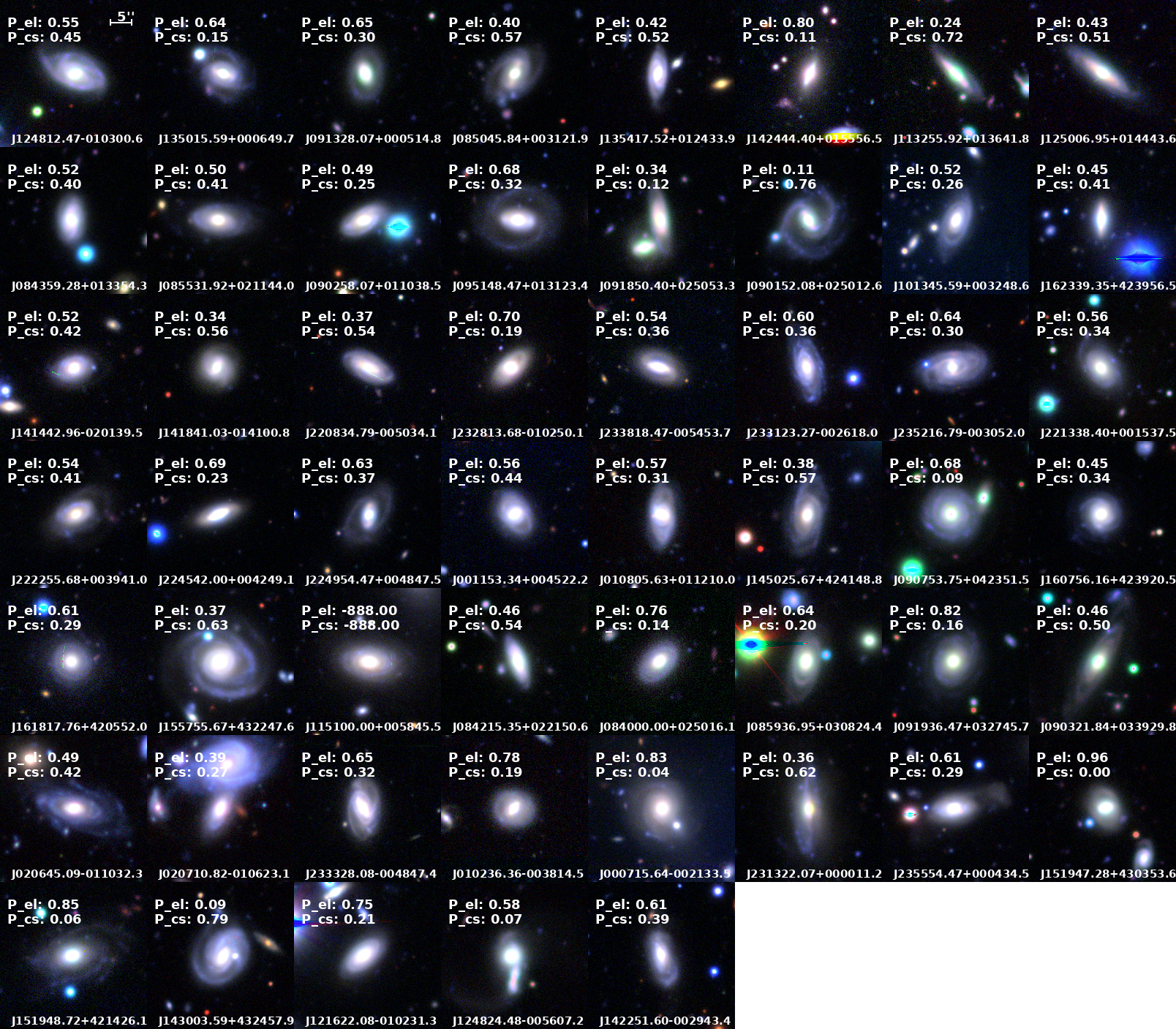}
\caption{\label{fig:collage}
Colour-composite HSC images in the $g$, $r$, and $i$ bands of 53 galaxies with observed-frame $g-r$ colour larger than $1.2$ and spiral morphology.
For each image the following morphological classification scores from Galaxy Zoo are shown: ${\rm P}_{\mathrm{el}}$ is the de-biased probability of the galaxy being elliptical, and ${\rm P}_{\mathrm{sp}}$ is the de-biased probability of it being a spiral.
}
\end{figure*}

{
For the purpose of gaining insight into the typical photometric properties of spiral arms, I obtained a sample of $53$ bulge-dominated spiral galaxies and carried out a decomposition of their light in a smooth surface brightness component and a set of spiral arms.
Section \ref{ssec:sample} introduces the sample of galaxies used for this analysis, and Sect. \ref{ssec:decompose} explains how the contribution of the spiral arms to the total surface brightness distribution was estimated.


\subsection{The sample}\label{ssec:sample}

I selected galaxies on the basis of their colour and morphology, as follows.
I considered objects from the main galaxy sample of the SDSS \citep{Str++02} as the starting point. 
I first defined a narrow redshift slice, $0.19 < z < 0.21$: this cut allows colours to be easily measured in a homogeneous way, without the need to apply K-corrections.
Then, I considered their observed-frame $g-r$ colours defined using `model' magnitudes from data release 16 of the SDSS \citep{Ahu++20} and applied the following cut:
\begin{equation}\label{eq:colorcut}
g - r > 1.2.
\end{equation}
Finally, I restricted the analysis to objects with available photometric data from the second public data release (PDR2) of the HSC survey \citep{Aih++19}.
I obtained $201\times201$ pixel cutouts (33.8'' per side) of background-subtracted, co-added images in the $g$, $r$, and $i$ bands from the HSC PDR2, made colour-composite images, and visually inspected them to select galaxies with spiral arms.
The resulting $53$ red spiral galaxies are shown in Fig. \ref{fig:collage}: these are the main subject of this section.
}

For the sake of comparing the HSC-based visual classification of this work with the literature, Fig. \ref{fig:collage} also shows morphological classification scores from Galaxy Zoo \citep{Lin++08, Lin++11}, which are based on SDSS images.
Interestingly, 39 out of the 53 galaxies have a higher probability of being elliptical than of being a spiral, according to Galaxy Zoo. This dramatic difference in the morphological classification of this set of objects is probably due to the lower depth and lower resolution of SDSS images,{ compared to HSC data}, which makes it difficult to see the spiral arms. 


\subsection{Photometric decomposition}\label{ssec:decompose}

I define as a spiral arm any positive excess flux that cannot be accounted for by a smooth model for the surface brightness distribution of a galaxy.{
In principle, this definition leads to the inclusion of features that are not spiral arms in a strict sense, such as rings or shells. In practice, as I discuss in Sect. \ref{sect:discuss}, the effect of such features on the fit of a S\'{e}rsic profile is qualitatively similar to that of proper spiral arms.
Therefore, I treat any flux excess associated with the main galaxy as part of a spiral arm.
}
With this definition, I proceeded as follows.
I fitted a S\'{e}rsic surface brightness profile to the HSC $r$-band images of the $53$ red spiral galaxies.
The $r$ band is the one that is most commonly used in photometric studies of galaxies with SDSS data.
The S\'{e}rsic model surface brightness is defined as follows:
\begin{equation}\label{eq:sersic}
I(x,y) = I_0\exp{\left\{-b(n)\left(\frac{R}{\reff}\right)^{1/n}\right\}},
\end{equation}
where $x$ and $y$ are Cartesian coordinates with origin at the galaxy centre and aligned with the galaxy major and minor axes, $R$ is the circularised radius
\begin{equation}\label{eq:circularised}
R^2 \equiv qx^2 + \frac{y^2}{q},
\end{equation}
$q$ is the axis ratio, $n$ is the S\'{e}rsic index, and $b(n)$ is a numerical constant defined in such a way that the isophote of radius $R=\reff$ encloses half of the total flux \citep[see][]{C+B99}.

\begin{figure*}
\includegraphics[width=\textwidth]{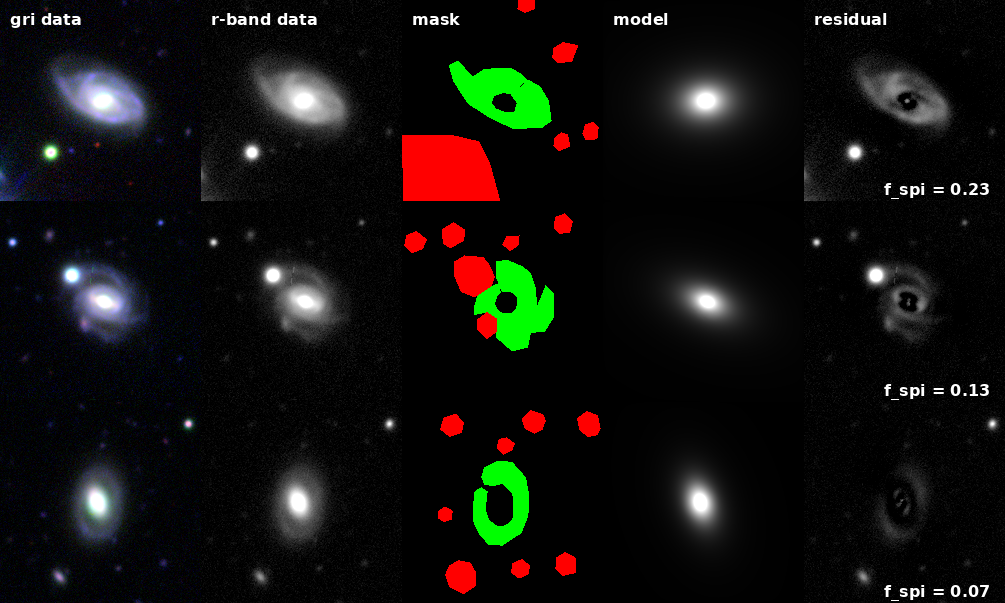}
\caption{
Fits of a S\'{e}rsic profile to the smooth component of the surface brightness distribution of three example galaxies.
Column 1: Colour-composite image of the galaxy in the HSC $g,r,$ and $i$ bands.
Column 2: Grey-scale image of the galaxy in the HSC $r$ band.
Column 3: Area over which the spiral arms dominate the surface brightness (green regions) and contaminants (red regions). Both are masked out from the fit.
Column 4: Best-fit model of the smooth component in the $r$ band.
Column 5: Residual in the $r$ band. The value of the $r$-band spiral-to-total flux ratio is shown in the bottom-right corner.
\label{fig:example}
}
\end{figure*}

This model is meant to describe only the smooth component of the surface brightness profile. For this reason, I manually masked out the region over which the spiral arms dominate the surface brightness when doing the fit.
Without doing so, the relative contributions of the smooth component and of the spiral arms to the total flux would be respectively overestimated and underestimated.
Examples of spiral arm masks are shown in the third column of Fig. \ref{fig:example}.

For each galaxy, I found the set of model parameters that minimises the following $\chi^2$:
\begin{equation}
\chi^2 = \sum_j \frac{\left(I_j^{\mathrm{(mod)}} - I_j^{\mathrm{(obs)}}\right)^2}{\sigma_j^2},
\end{equation}
where $I_j^{\mathrm{(mod)}}$ is the seeing-convolved S\'{e}rsic model in the $r$ band evaluated at pixel $j$, $I_j^{(\mathrm{obs})}$ is the observed surface brightness at that pixel, and $\sigma_j$ is the observational uncertainty.
I assumed flat priors on all of the model parameters.
For the S\'{e}rsic index, I imposed lower and upper limits at $0.5 < n < 8$.
I sampled the parameter space of the structural parameters with a Markov chain Monte Carlo. At each step of the chain I found the value of the amplitude of the model (parameter $I_0$ in Eq. \ref{eq:sersic}) that minimises the $\chi^2$.
Figure \ref{fig:example} shows fits for three example galaxies.

\begin{figure*}
\includegraphics[width=\textwidth]{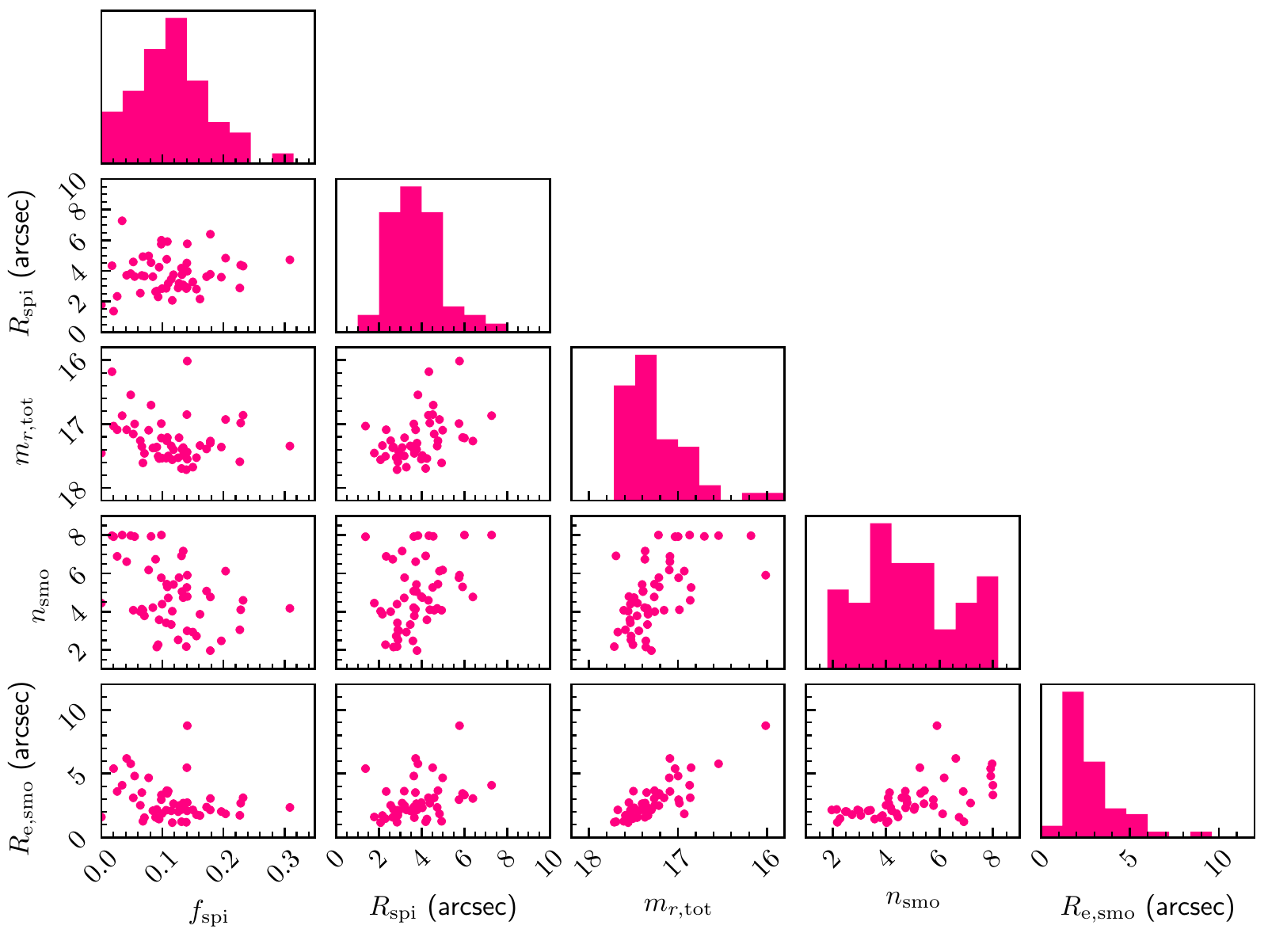}
\caption{
Distribution in spiral-arm-to-total ratio, $\fspi$, spiral arm light-weighted radius, $\Rspilw$, total r-band magnitude, $m_{r,\mathrm{tot}}$, half-light radius of the smooth component, $R_{\mathrm{e,smo}}$, and S\'{e}rsic index of the smooth component, $n_{\mathrm{smo}}$, of the 53 red spiral galaxies in the sample.
\label{fig:cp}
}
\end{figure*}

After finding the minimum-$\chi^2$ set of parameters, I subtracted the best-fit model from the data. I then estimated the spiral arm flux by taking the sum of the residual flux in the masked region.
I summarised the properties of the spiral arms with two numbers.
The first is the $r$-band spiral-arm-to-total ratio, $\fspi$, defined as the ratio between the flux associated with the spiral arms and the total flux.
The second quantity is the light-weighted radius, defined as
\begin{equation}
\Rspilw = \dfrac{\sum_{i\in{\mathrm{spi}}} I_i\cdot R_i}{\sum_{i\in{\mathrm{spi}}} I_i},
\end{equation}
where $I_i$ is the surface brightness of the spiral arm at the $i$-th pixel, $R_i$ is the circularised radius, defined with respect to the smooth component, at the same pixel, and the sum is taken over all pixels in the spiral arm mask with non-negative flux.
Figure \ref{fig:cp} shows the distribution in $\fspi$, $\Rspilw$, total (smooth component plus spiral arms) $r$-band magnitude, and half-light radius and S\'{e}rsic index of the smooth component.

The smooth component covers a broad range in values of the S\'{e}rsic index, with most of the galaxies having $n_{\mathrm{smo}} > 4$, which is typical of bulges.
The median value of the spiral-arm-to-total ratio is $\fspi = 0.11$, while the median value of the ratio between the light-weighted radius and the half-light radius of the smooth component, $\Rspilw/R_{\mathrm{e,smo}}$, is approximately $1.5$.


\section{Tests on simulated galaxies}\label{sect:results}

In this section I quantify how the presence of spiral arms affects the accuracy of the inference of the total flux and half-light radius of a galaxy when these are measured by means of the fit of a S\'{e}rsic model. 
This problem is a high-dimensional one:
the answer can depend, in principle, on many different features of the surface brightness distribution of the galaxy, but also on the details of the data used for the fit, as well as on aspects of the fitting procedure (for instance, the choice of priors).
In this work, I focus on the following properties: the spiral-arm-to-total ratio, $\fspi$, the ratio between the light-weighted radius of the spiral arm and the half-light radius of the smooth surface brightness component, $\Rspilw/R_{\mathrm{e,smo}}$, the S\'{e}rsic index of the smooth component, and the depth of the photometric data.

In order to isolate the effect that each of these aspects has on the inference, I simulated different sets of simulated images, varying only one property at a time.
Section \ref{ssec:generalsim} describes the general characteristics of the simulations. The ensuing sections describe each set of simulations and the results obtained by fitting a S\'{e}rsic model to them.

\subsection{Simulations. General characteristics}\label{ssec:generalsim}

Each simulated galaxy consists of a smooth surface brightness component, described by a S\'{e}rsic profile, and a set of spiral arms.
Throughout the experiment, I fixed the half-light radius of the smooth component to $R_{\mathrm{e,smo}} = 2.52''$ ($15$ HSC pixels), its $r$-band magnitude to $m_{r,\mathrm{smo}} = 17.5$, and its axis ratio to one.

The shape of the spiral arms is described by the following function \citep{R+M09}:
\begin{equation}\label{eq:spiralarm}
\Rspi(\phi) = -\frac{A}{\ln{\left(B\,\rm{tan}\frac{\phi}{2N}\right)}},
\end{equation}
where $R$ and $\phi$ are polar coordinates, and $A$, $B,$ and $N$ are free parameters.
This model describes a single spiral arm originating at angle $\phi=0$. A second spiral arm can be obtained by rotating the coordinate frame by $\pi$.
The arm extends along positive values of $\phi$ as long as both of the following conditions are met:
\begin{align}
B\,\rm{tan}\frac{\phi}{2N} < 1, \\
R(\phi) < R_{\mathrm{max}},
\end{align}
where $R_{\mathrm{max}}$ is a parameter describing the maximum allowed distance from the galaxy.
The first condition ensures that the logarithm in Eq. \ref{eq:spiralarm} returns a negative value, so that $R$ is positive, while the second condition is added in order to avoid producing excessively long spiral arms.
For all simulated galaxies, I fixed the value of $R_{\mathrm{max}}$ to $0.5A$.

\Eref{eq:spiralarm} describes a spiral arm with a pitch angle\footnote{The pitch angle of a spiral arm is the angle between its tangent and a circle centred on the galaxy centre.} that decreases as a function of radius.
The parameter $A$ sets the scale of the arm, while the parameters $B$ and $N$ affect how rapidly the pitch angle varies as a function of $\phi$.

\Eref{eq:spiralarm} defines the position of the core of each arm.
I then assumed that the contribution of the arm to the surface brightness distribution of the galaxy is given by the following exponential function,
\begin{equation}
I_{\mathrm{spi}}(R,\phi) = I_{\mathrm{spi},0} \exp{\left\{-\frac{|R - \Rspi(\phi)|}{h}\right\}},
\end{equation}
where the parameter $h$ sets the width of the arm. I fixed the value of $h$ to two HSC pixels (0.336'') for all cases.

Given a set of model parameters that describe a spiral galaxy, I generated HSC-like images by convolving the model surface brightness distribution with the HSC point spread function. I added to each pixel a Gaussian noise with dispersion $\sigma_{\mathrm{sky}}$ equal to that observed in HSC data, as well as Poisson noise, assuming an exposure time of $10$~minutes (the median exposure time of the HSC PDR2 in the $r$ band).
As in Sect. \ref{sect:spifrac}, the analysis is based on $r$-band images.
Nevertheless, for the purpose of creating colour-composite images, I generated data in the $g$- and $i$-band data as well.
In all simulations, the colours of the smooth component are fixed to $g-r = 1.2$ and $r-i = 0.5$, while those of the spiral arms are fixed to $g-r = 0.5$ and $r-i=0.1$.

I generated four sets of simulated galaxies, each aimed at testing the effect of a given property on the bias on the S\'{e}rsic photometry. The characteristics of each set are summarised in \Tref{tab:sims}.
\begin{table*}
\caption{Simulation properties}
\label{tab:sims}
\begin{tabular}{c|cccccc}
\hline
\hline
Set & $\fspi$ & $n_{\mathrm{smo}}$ & $A$ & $B$ & $N/B$ & $\sigma_{\mathrm{sky}}$ \\
\hline
A & $0.1$ & $4$ & $[4'', 8'', 12'', 16'', 20'']$ & $[0.1, 1]$ & $[5, 10, 20]$ & $\sigma_{\mathrm{HSC}}$ \\
B & $0.1$ & $[1, 3, 5]$ & $[4'', 8'', 12'', 16'', 20'']$ & $1$ & $10$ & $\sigma_{\mathrm{HSC}}$ \\
C & $[0.01, 0.05, 0.1, 0.2]$ & $4$ & $12''$ & $1$ & $10$ & $\sigma_{\mathrm{HSC}}$ \\
D & $0.1$ & $4$ & $12''$ & $1$ & $10$ & $[0.2, 0.5, 1, 2, 10, 20, 50]\times\sigma_{\mathrm{HSC}}$
\end{tabular}
\tablefoot{Simulation parameters for each set. Column (2): Spiral-arm-to-total flux ratio in the $r$ band. Column (3): S\'{e}rsic index of the smooth component. Column (4): Scale parameter of the spiral arms. Column (5): Parameter $B$ of the spiral arms. Column (6): Ratio $N/B$ of the spiral arms. Column (7): Sky background fluctuation level. The quantity $\sigma_{\mathrm{HSC}}$ indicates the sky background noise level of HSC data.}
\end{table*}

\subsection{Set A. Effect of the shape and position of the spiral arms}

\begin{figure*}
\includegraphics[width=\textwidth]{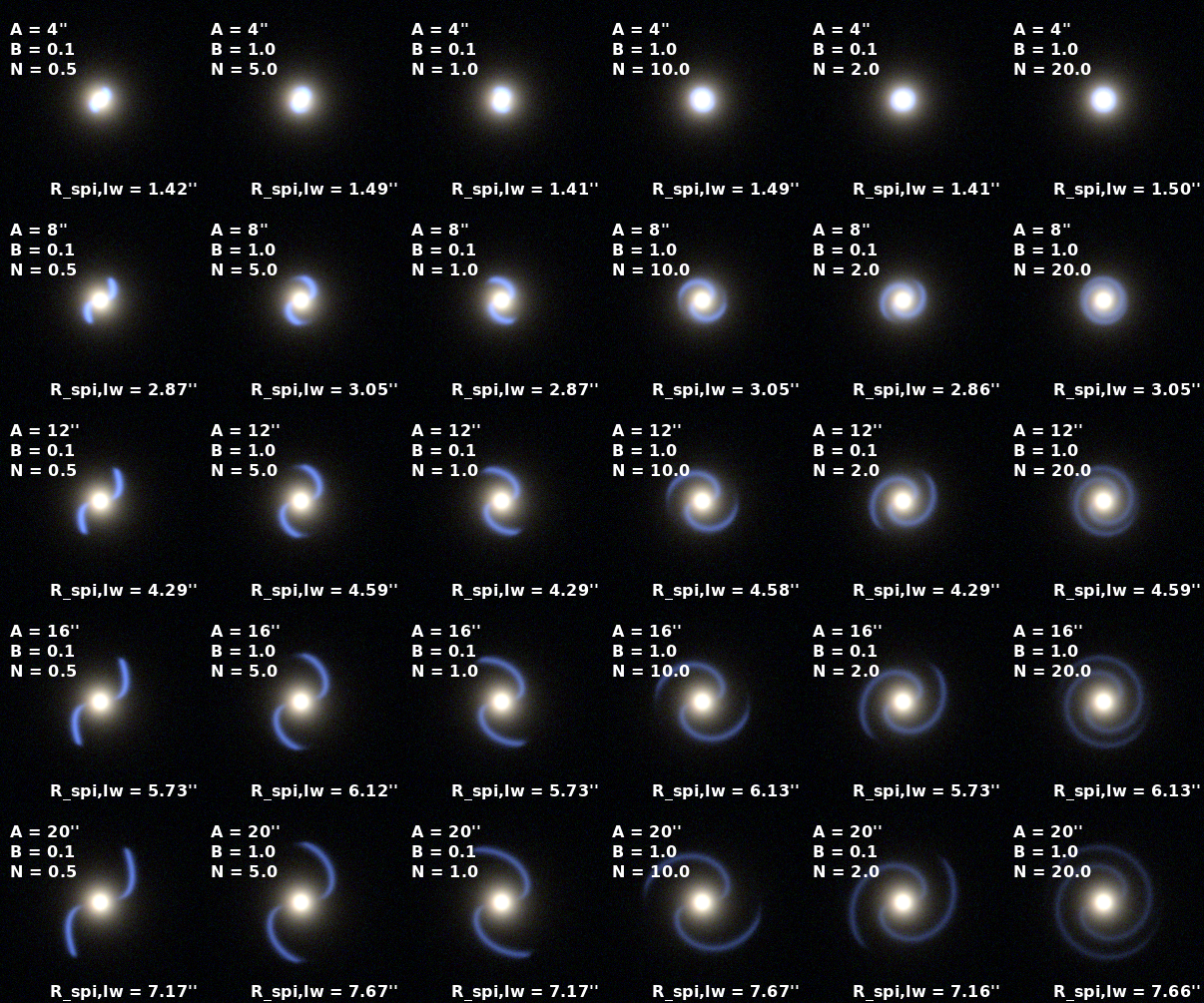}
\caption{
Colour-composite images in HSC $g,r,i$ of red spiral galaxies from simulation set A.
Simulations consist of the sum of a smooth component, a S\'{e}rsic profile with $n=4$ and $\reff = 2.52''$, and a set of spiral arms, accounting for $10\%$ of the total $r$-band flux. The spiral arm parameters cover a grid of values, as indicated in the panels.
The values of the $r$-band half-light radius and the light-weighted radius of the spiral arm are shown in the bottom right of each panel.
\label{fig:sims}
}
\end{figure*}

The first set of simulations, labelled as set A, consists of $30$ galaxies with a fixed smooth component and varying properties of the spiral arms. Its purpose is to check the effect that the shape and the position of the spiral arms has on the inferred flux and size of the galaxy.
The S\'{e}rsic index of the smooth component is fixed to $n_{\mathrm{smo}}=4$. 
The spiral arm parameters cover a grid of five values of $A$, two values of $B,$ and three values of the ratio $N/B$, as indicated in the first row of \Tref{tab:sims}.
In all cases, the $r$-band spiral-arm-to-total ratio is fixed to $\fspi=0.1$.
The value of the resulting total $r$-band magnitude is $m_{r,\mathrm{tot}} = 17.40$. 
The half-light radius, which I define here as the radius of the circle that encloses half of the total flux in the $r$ band, varies depending on the location of the spiral arms: it increases with increasing $\Rspilw$. 
Figure \ref{fig:sims} shows colour-composite images of these simulated galaxies and reports the values of $\Rspilw$ and $\reff$ of each galaxy.

\begin{figure}
\includegraphics[width=\columnwidth]{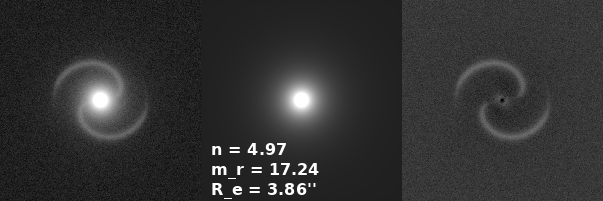}\\
\caption{
Fit of a S\'{e}rsic profile to the $r$-band image of a simulated galaxy.
Left panel: $r$-band image. Middle panel: Image of the minimum-$\chi^2$ model. Right panel: Residual.
The galaxy is taken from simulation set A, with spiral arm parameters $A=16''$, $B=1$, $N=10,$ and $\fspi=0.1$ and smooth component parameters $n_{\mathrm{smo}} = 4$, $\reff=2.52''$, and $m_{r,\mathrm{smo}} = 17.5$. The true values of the total $r$-band magnitude and half-light radius are $m_{r,\mathrm{tot}} = 17.40$ and $R_{\mathrm{e}} = 3.06''$.
The parameters of the best-fit S\'{e}rsic model are reported in the top-middle panel.
\label{fig:example_simfit}
}
\end{figure}

I fitted the $r$-band image of each simulated galaxy with a S\'{e}rsic profile using the same procedure described in Sect. \ref{sect:spifrac} but without masking the spiral arms.
Figure \ref{fig:example_simfit} shows the results of the fit in an example case.
Both the total flux and the half-light radius are overestimated.
The residuals show over-subtraction in the region enclosed within the spiral arms and positive flux on the spiral arms themselves.

\begin{figure}
\includegraphics[width=\columnwidth]{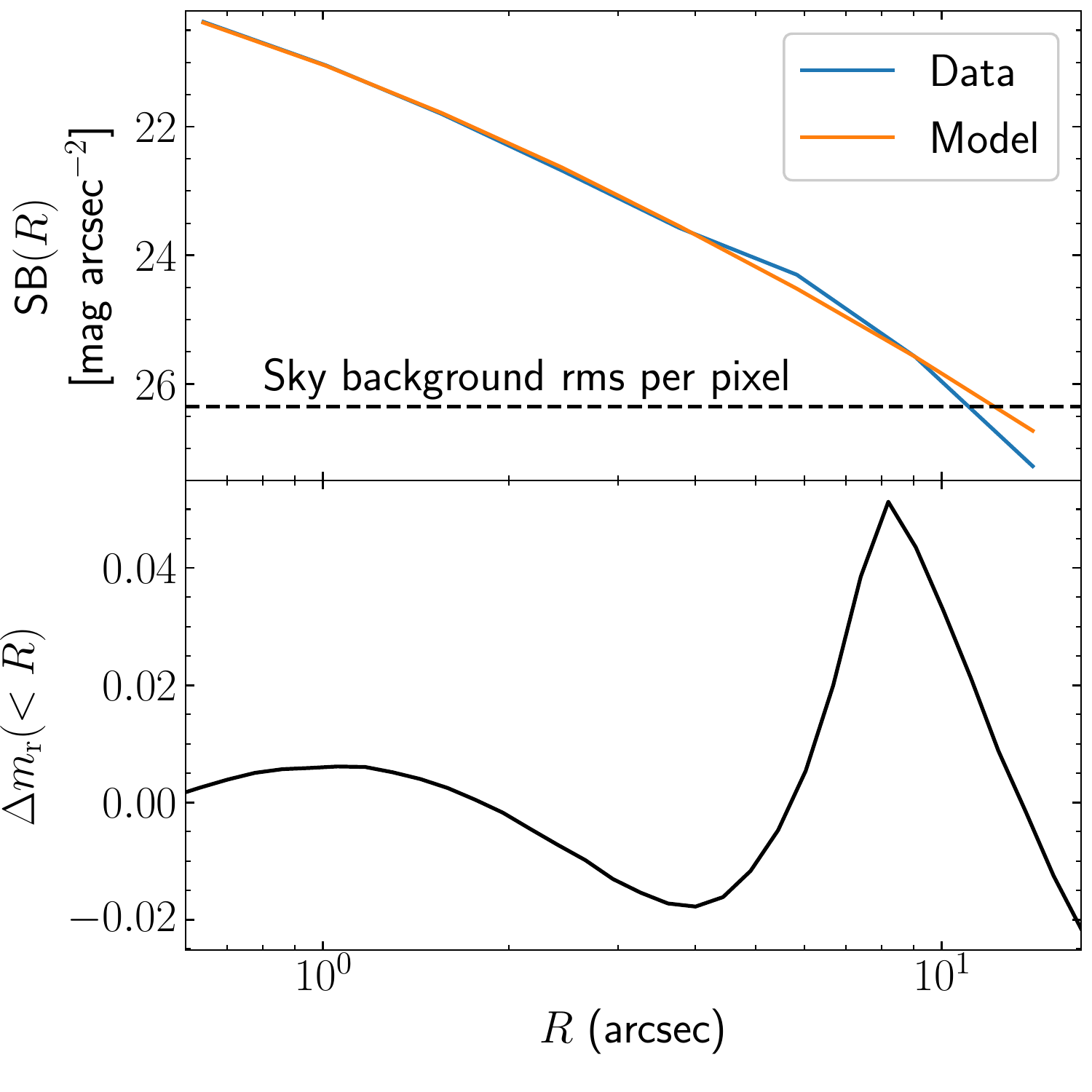}
\caption{
Quantitative analysis of the fit of \Fref{fig:example_simfit}.
Top: Azimuthally averaged surface brightness profile of the galaxy and best-fit S\'{e}rsic model.
The horizontal dashed line marks the sky background root mean square surface brightness fluctuation for a single pixel.
Bottom: Difference between the magnitude within radius $R$ of the model and that of the true galaxy image, as a function of $R$.
\label{fig:fit1d}
}
\end{figure}

Figure \ref{fig:fit1d} offers a more quantitative view of this result, as it shows the azimuthally averaged surface brightness profiles of the simulated galaxy and of the best-fit S\'{e}rsic model, as well as the difference in the enclosed flux profile between the model and the truth.
The spiral arms create a bump in the surface brightness profile at $\approx6''$.
The model is unable to fit both this bump and the sharp decline in surface brightness at larger radii.
As a result, the best-fit model has a shallower surface brightness profile than the truth at large radii.
Although the bias in the integrated flux is small within the region constrained by the data (only $0.02$~mag for $R<15''$, as shown by the bottom panel of Fig. \ref{fig:fit1d}), the extrapolation to the large radii of the model results in biases of $0.16$~mag on the total flux and $26\%$ on the half-light radius.

\begin{figure}
\includegraphics[width=\columnwidth]{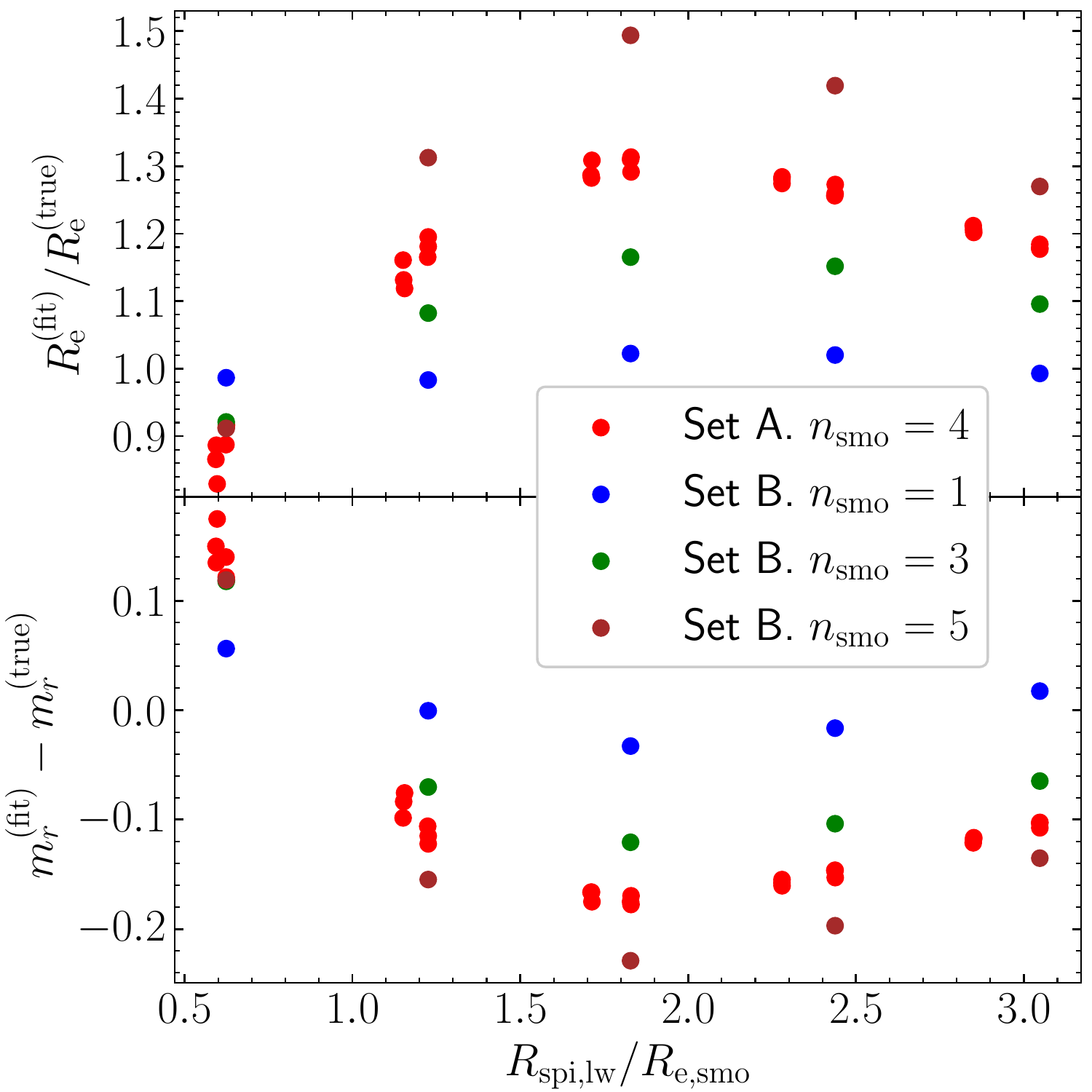}
\caption{
Bias on the $r$-band magnitude (bottom panel) and half-light radius (top panel) from the fit of a S\'{e}rsic profile to the $r$-band images of the set of simulated galaxies shown in Fig. \ref{fig:sims}, as a function of the ratio between the light-weighted radius of the spiral arms and the half-light radius of the smooth surface brightness component. 
All of the simulated galaxies have a spiral-arm-to-total ratio of $\fspi=0.10$ and a smooth component with $n=4$ and $R_{\mathrm{e,smo}}=2.52''$, and the images have HSC-like properties.
\label{fig:rtrend}
}
\end{figure}

Figure \ref{fig:rtrend} shows the bias on the inferred $r$-band magnitude and half-light radius for simulated set A (red points). These biases are plotted as a function of the ratio between the light-weighted radius of the spiral arms and the half-light radius of the smooth component, $\Rspilw/R_{\mathrm{e,smo}}$.
We see a very clear trend between $\Rspilw/R_{\mathrm{e,smo}}$ and the bias. 
Flux and size are underestimated in galaxies with very small spiral arm radii and overestimated for $\Rspilw/R_{\mathrm{e,smo}} > 1$.
Remarkably, there is very little scatter around the mean trend: for this simulated set, the bias appears to be entirely described by the position of the spiral arms, while the shape of the arms themselves only produces minimal deviations.
The maximum bias occurs at a value of $\Rspilw/R_{\mathrm{e,smo}}$ between $1.5$ and $2.0$. Interestingly, this is close to the median value of the real sample of red spirals measured in Sect. \ref{sect:spifrac}.
For galaxies of this kind, the inferred magnitude is smaller than the true value by $0.18$, which is equivalent to the flux being overestimated by $0.06$~dex (15\%). The half-light radius is overestimated by $30\%$.

\subsection{Set B. Effect of the S\'{e}rsic index of the smooth component}

Simulation set B consists of 15 galaxies generated by varying the S\'{e}rsic index of the smooth component, $n_{\mathrm{smo}}$, and the scale parameter, $A$, of the spiral arms, as described in \Tref{tab:sims}.
The goal of these simulations is to check to what extent the results obtained in the previous section depend on the choice of $n_{\mathrm{smo}}$.
The results of the S\'{e}rsic profile fit are shown in Fig. \ref{fig:rtrend}.

At fixed $n_{\mathrm{smo}}$, galaxies from this set also show a trend between the bias and the ratio $\Rspilw/R_{\mathrm{e,smo}}$, with the maximum bias occurring at $\Rspilw/R_{\mathrm{e,smo}} \approx 1.8$.
The amplitude of the bias, however, appears to be a strong function of the S\'{e}rsic index of the smooth surface brightness component.
While galaxies with $n_{\mathrm{smo}}=5$ can have a bias on the total magnitude as large as $0.24$~mag, the largest bias observed in galaxies with $n_{\mathrm{smo}}=1$ and $\Rspilw/R_{\mathrm{e,smo}} > 1$ is only $0.04$~mag.
Simulations with smaller values of the ratio $\Rspilw/R_{\mathrm{e,smo}}$ can produce larger biases, but such galaxies are not very realistic.

The reason for this difference lies in the large-scale behaviour of the surface brightness profile as a function of $n_{\mathrm{smo}}$.
The higher the S\'{e}rsic index, the larger the fraction of the total flux that is associated with the outer regions of the galaxy.
For example, for a S\'{e}rsic profile with $n=1$, a circular aperture with radius equal to two half-light radii encloses $85\%$ of the total flux.
The same aperture for a S\'{e}rsic profile with $n=4$, however, encloses only 69\% of the total flux.
The presence of the spiral arms causes the best-fit S\'{e}rsic model to deviate from the surface brightness profile of the smooth component at large radii (as shown in Fig. \ref{fig:fit1d}).
If the value of $n_{\mathrm{smo}}$ is small, this deviation does not introduce a large bias, because most of the total flux is enclosed in a region that is well constrained by the data.
If $n_{\mathrm{smo}}$ is large, however, differences in the large-scale behaviour between the true surface brightness profile and the model can have a higher impact.

In the case shown in Fig. \ref{fig:fit1d}, for example, the bias on the flux enclosed within an aperture of $16''$ (the size of the cutout) is as small as $0.02$~mag.
However, since the best-fit S\'{e}rsic profile has an index $n=4.97$ and a half-light radius of $\reff = 3.86''$, regions at $R>16''$ still contribute 17\% of the model total flux (the difference between the total magnitude and the $16''$ aperture magnitude is $0.20$), and the extrapolation to large radii can therefore introduce a larger bias.
Increasing the size of the cutout does not offer any help in this case, because the surface brightness of the galaxy already falls below the sky background fluctuation level at $R\approx10''$.

\subsection{Set C. Trends with $f_{\mathrm{spi}}$}

\begin{figure}
\includegraphics[width=\columnwidth]{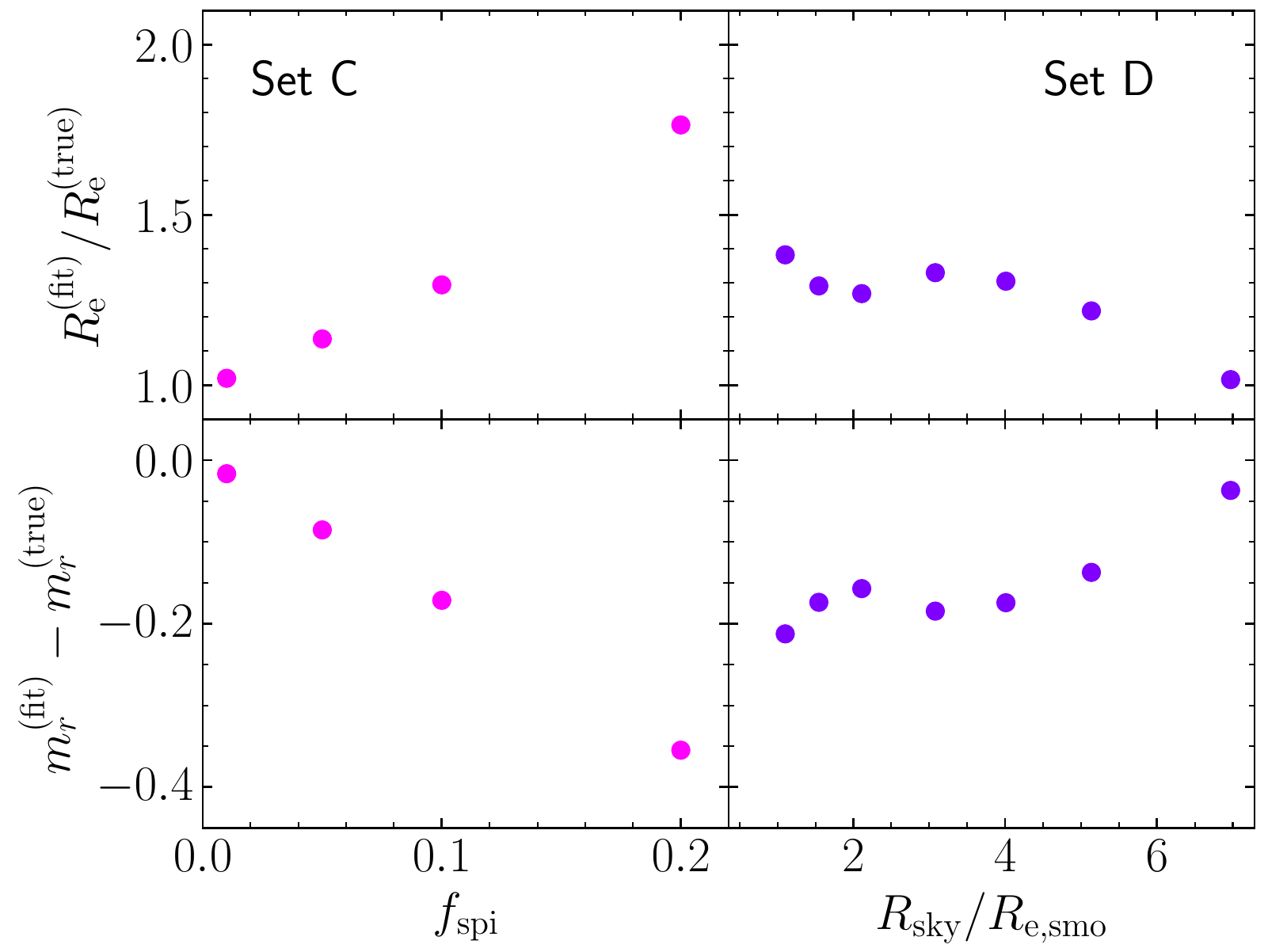}
\caption{
Bias on simulation sets C and D.
Left: Bias on the half-light radius (top) and total magnitude (bottom) as a function of the spiral-arm-to-total flux ratio, obtained by fitting a S\'{e}rsic profile to images from simulation set C.
Right: Bias on half-light radius (top) and total magnitude (bottom) as a function of the ratio $R_{\mathrm{sky}}/R_{\mathrm{e,smo}}$ for simulation set D. The quantity $R_{\mathrm{sky}}$ is defined as the radius at which the surface brightness of the smooth component equals the amplitude of the sky background fluctuation for a single pixel.
\label{fig:setCD}
}
\end{figure}

Simulation set C consists of four galaxies with a fixed smooth component, fixed spiral arm shape parameters, and different values of the spiral-arm-to-total flux ratio, $\fspi$.
Its purpose is to understand how the amplitude of the bias scales with $\fspi$.
The results of the S\'{e}rsic fits to these galaxies are shown in the left-hand column of Fig. \ref{fig:setCD}.
When $\fspi=0.01$, the bias is very small: this is a sanity check, showing that, in the limit in which the spiral arms are very faint, the S\'{e}rsic model is able to recover the true magnitude and size of the galaxy (which is dominated by the smooth component).
The other data points show that the amplitude of the bias increases approximately linearly with $\fspi$.
%

\subsection{Set D. Effect of the depth of the data}\label{ssec:depth}

The last test is aimed at checking how the bias depends on the depth of the data used for the fit. 
For this reason, simulation set D consists of images of the same galaxy that differ only by the amplitude of the background noise.
In particular, indicating with $\sigma_{\mathrm{HSC}}$ the $r$-band sky background fluctuation of the HSC data, I generated images with sky background fluctuation amplitude equal to $\sigma_{\mathrm{sky}} = f \sigma_{\mathrm{HSC}}$, where $f$ varies from $0.2$ (highest depth) to $50$ (lowest depth).
The results of fits to these images are shown in the right-hand column of Fig. \ref{fig:setCD}.
The bias on the magnitude and half-light radius are plotted as a function of the quantity $R_{\mathrm{sky}}/R_{\mathrm{e,smo}}$, where $R_{\mathrm{sky}}$ is the radius at which the surface brightness of the smooth component is equal to the amplitude of the background noise fluctuation for a single pixel. 
Roughly speaking, the galaxy is detected on pixels that lie within a circle of radius $R_{\mathrm{sky}}$ and are undetected outside it. 
Higher values of $R_{\mathrm{sky}}$, then, correspond to deeper data.
Figure \ref{fig:setCD} shows that the accuracy of the fit improves with increasing depth of the data: as information on the outer profile of a galaxy becomes available, the model is able to better recover its total flux.


\section{Discussion and summary}\label{sect:discuss}

The experiment carried out in Sect. \ref{sect:results} shows that the presence of spiral arms can cause a bias in the determination of the total flux and half-light radius of a galaxy when these are determined by fitting a S\'{e}rsic profile to its surface brightness distribution.
Given a galaxy consisting of the sum of a smooth surface brightness component, following a S\'{e}rsic profile, and a set of spiral arms, the amplitude of the bias is most sensitive to the relative contribution of the arms to the total flux, the position of the arms relative to the half-light radius of the smooth component, the S\'{e}rsic index of the smooth component, and the depth of the data.
In particular, for a $z=0.2$ galaxy with a S\'{e}rsic index of the smooth component $n_{\mathrm{smo}}=4$, a set of spiral arms accounting for 10\% of the total flux (which is typical, according to the analysis of Sect. \ref{sect:spifrac}) causes the total flux to be overestimated by $0.06$~dex (15\%) and the half-light radius to be overestimated by $30\%$.
Conversely, the bias is very small if the smooth component of the surface brightness profile has a S\'{e}rsic index close to $1$.

\begin{figure}
\includegraphics[width=\columnwidth]{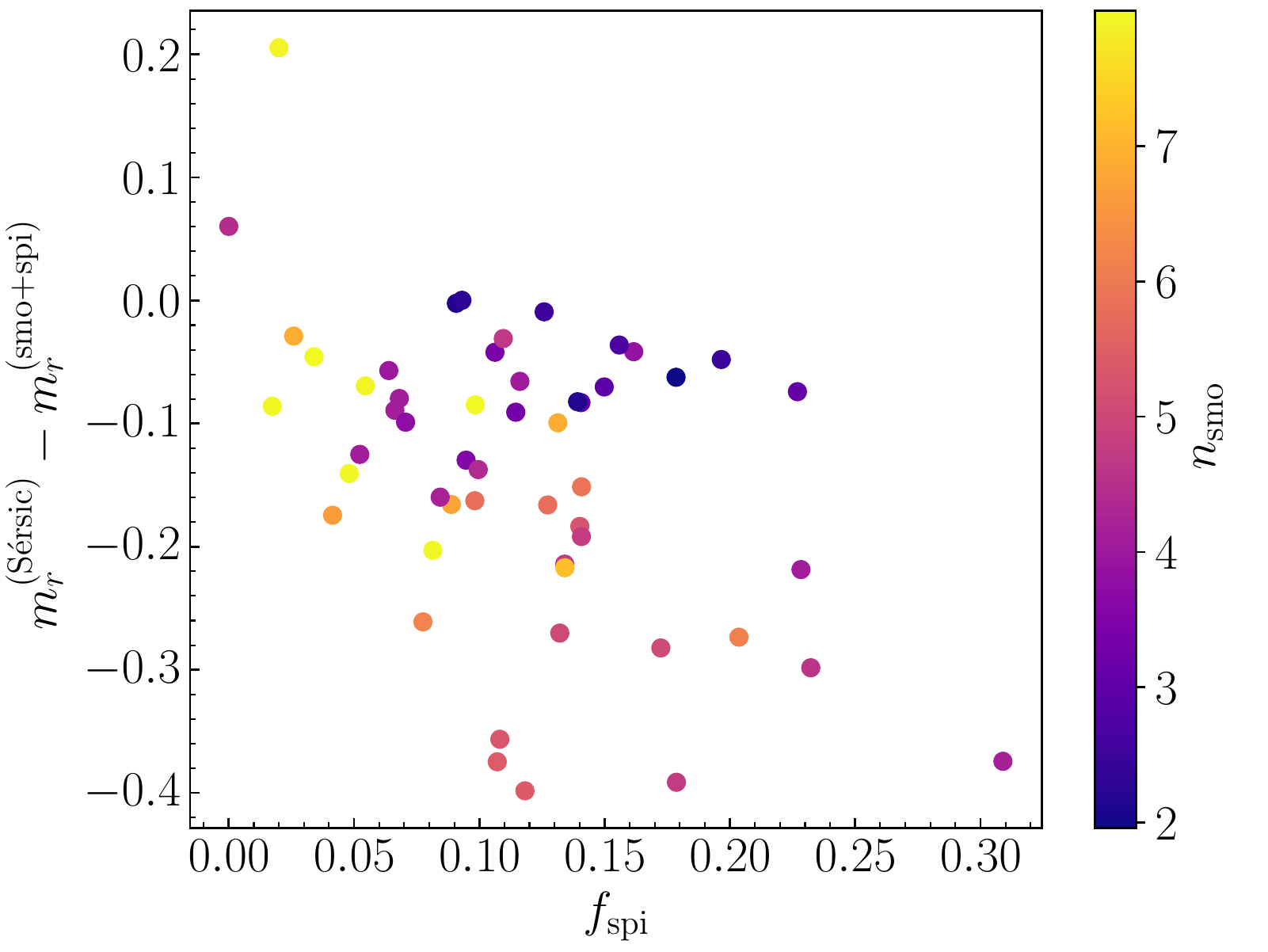}
\caption{
Difference in the $r$-band magnitude of the $53$ red spiral galaxies of Sect. \ref{sect:spifrac} obtained by fitting a S\'{e}rsic profile and with a two-component description (smooth S\'{e}rsic component plus spiral arms) as a function of the spiral arm fraction, $\fspi$.
The points are colour-coded by the S\'{e}rsic index of the smooth component.
\label{fig:realsample}
}
\end{figure}

These results were obtained on simulated images and, strictly speaking, only apply to galaxies with properties that match exactly those of the simulations. 
In order to make sure that they can be generalised to real cases, I carried out the following test.
I took the 53 red spiral galaxies introduced in Sect. \ref{ssec:sample} and compared the total $r$-band flux measured in Sect. \ref{ssec:decompose}, which was obtained by separating out the spiral arms and the smooth component, with that obtained by fitting a S\'{e}rsic profile to the full data (without masking the spiral arms).
Figure \ref{fig:realsample} shows the difference in the total magnitude obtained with these two methods as a function of the $r$-band spiral arm fraction, $\fspi$.

For all galaxies but two, the single S\'{e}rsic profile fit produces $r$-band fluxes that are larger than that those obtained with the two-component description.
The difference correlates with the spiral arm fraction and also with the S\'{e}rsic index, qualitatively confirming the findings of Sect. \ref{sect:results}.
The median difference in magnitude between the two-component and single S\'{e}rsic fit is $-0.11$~mag.
This test shows that real galaxies behave in a similar way to the ones simulated in Sect. \ref{sect:results} with respect to the effect of spiral arms on the fit of S\'{e}rsic fits.
It does not imply, however, that the fluxes obtained with the two-component fit are perfectly accurate: these measurements are based on the assumption of a S\'{e}rsic profile for the smooth component and therefore could still be biased if this assumption breaks down at large radii. 
This is particularly important for fits with a large S\'{e}rsic index of the smooth component, since a relatively large fraction of their total flux is due to the extrapolation of the profile to large radii.
As I discuss below, the problem of determining accurate total fluxes of spiral galaxies is an open one.

In light of these results, one can expect that the galaxies that are most susceptible to biases are those with a dominant bulge component, that is, a smooth surface brightness profile with a large ($n > 2.5$) S\'{e}rsic index, such as the red spiral galaxies studied in the first part of this paper.
Red spiral galaxies are relatively rare: they make up only 14\% of the total of red galaxies in the sample of Sect. \ref{sect:spifrac}.
However, the population of spiral galaxies with a prominent bulge extends well into the blue galaxy cloud.
Of the SDSS main sample galaxies with $g-r < 1.2,$  40\% have a S\'{e}rsic index $n>2.5$, as measured by \citet{MVB15}. Upon visual inspection, 60\% of them have spiral arms (24\% of the total of blue galaxies).

In general, this bias is more important for samples of galaxies in which there is a comparable mixture of disks and bulges. 
At redshift zero, this occurs roughly at stellar masses $\log{M_*/M_\odot} \sim 10.5$-$11.0$, also known as the knee of the galaxy stellar mass function.
The overlap between the late-type and early-type galaxy populations at this stellar mass is well illustrated by \citet{Tay++20} in their Fig. 1, which shows the distribution in various properties of a large sample of galaxies in a narrow stellar mass range around $\log{M_*/M_\odot}\sim10.5$.
Their data show the well-known bimodality in colour, with red and blue galaxies accounting for similar fractions of the full sample.
Interestingly, the blue galaxy population spans a broad range in S\'{e}rsic index, with approximately 20\% of them above the limit of $n=2.5$ that is often used to separate disks from bulges.

The spiral arm bias could have an impact on measurements aimed at detecting differences between the red and blue galaxy populations at fixed stellar mass, such as the studies of the average dark matter halo mass of these two galaxy types \citep[for example, ][]{Mor++11,Zeh++11,Tin++13,Hud++15,Man++16}. 
If the total flux is overestimated for a set of galaxies, so is their stellar mass. Since halo mass correlates strongly with stellar mass, that would lead to a biased interpretation of the results.
Assuming that the test of Fig. \ref{fig:realsample} is representative of the general galaxy population, then the typical bias due to spiral arms is $\sim0.1$~mag on the total flux.
Based on the arguments made above in this section, one can expect that only 20\% of the galaxies, those with both a prominent bulge and spiral arms, are affected by this bias. 
Therefore, given the current observational uncertainties of the stellar-to-halo mass relation, the bias is probably not significant.
Nevertheless, systematic errors related to the photometric measurements can become important as precision improves.

This work focused only on the effect of the spiral arms on the results of S\'{e}rsic profile fits: throughout the paper, the bias was defined with respect to the ideal case in which the true surface brightness profile of a galaxy consists of a smooth component that follows a S\'{e}rsic profile and a set of spiral arms.
If the true surface brightness profile of the smooth component of a galaxy deviates from a S\'{e}rsic model, however, the measurement will suffer from additional biases.
The tests carried out in this work, therefore, represent a best-case scenario.
{ 
Additionally, real spiral galaxies often show ring-like features (see for example the third row in Fig. \ref{fig:example}), which are not directly captured by the model of Eq. \ref{eq:spiralarm}.
Nevertheless, I carried out tests on simulated galaxies consisting of the sum of a S\'{e}rsic and a ring-like component, using the ring model of \citet{Son++18}, and found very similar results to those shown in Sect. \ref{sect:results}, at fixed $\fspi$.
}

The problem of determining the total flux of a galaxy is a difficult one.
The results of Sect. \ref{ssec:depth} show that increasing the depth of the data can help greatly in reducing the spiral arm bias.
However, it is important to point out that all of the other tests carried out in this work are based on galaxies at relatively low redshift, $z=0.2$, observed with HSC-like data. 
While upcoming surveys such as the Legacy Survey of Space and Time (LSST\footnote{\url{https://www.lsst.org/}}) will go approximately one magnitude deeper than HSC, this will still be insufficient to probe the outskirts of higher redshift galaxies.

A method that is often adopted to fit galaxies with complex morphology is that of performing a bulge-disk decomposition. 
This is usually done by describing the total surface brightness as the sum of a de Vaucouleurs profile (a S\'{e}rsic profile with $n=4$) for the bulge and an exponential profile ($n=1$) for the disk.
Unfortunately, there are limitations with this approach, as recently pointed out by \citet{Pap++22}. The disk is generally associated with the younger component of the stellar population of a galaxy. This young component, however, is often missing in the inner regions of galaxies, where the star formation activity has stopped, causing the true surface brightness profile to deviate from a simple exponential model.
Moreover, the assumption of a de Vaucouleurs profile for the bulge component might also lead to biases if the true surface brightness profile deviates from an $n=4$ model at large radii.

A better solution would be to increase the complexity of the model used to describe a galaxy.
Although there exists software that allows complex spiral arm structure to be fit \citep[for example {\sc Galfit}; ][]{Pen++02}, this approach typically requires human interaction and does not scale well to very large samples of galaxies.
A powerful alternative is the use of neural networks trained on a large set of simulated galaxy images \citep[see for example][]{Smi++21, Li++21}.
While neural networks are very efficient at producing fits to galaxies with complex morphology, their output depends critically on the simulations used to train them. 
If the training set images are biased with respect to the true surface brightness distribution of galaxies, for instance at large radii, then they will introduce biases on the total flux.

Given how challenging it is to accurately estimate the total flux of a galaxy, an alternative is to instead quantify galaxy fluxes over finite apertures.
One example is the description in terms of the total flux within $10$~kpc and light-weighted surface brightness slope in the same aperture, introduced by \citet{Son20} for elliptical galaxies.
As the bottom panel of Fig. \ref{fig:fit1d} shows, the S\'{e}rsic model is able to correctly measure the total flux within a finite aperture with a precision of $\sim0.02$~mag.
This precision could be further improved with the use of non-parametric fitting methods, such as the isophote-fitting tool {\sc AutoProf} \citep{Sto++21}, or the segmentation map-based software {\sc PROFOUND} \citep{Rob++18}.

In conclusion, while the S\'{e}rsic profile is a very useful model for approximating the surface brightness distribution of a large set of galaxies, it can introduce systematic errors on the total flux and half-light radius of galaxies with spiral arms.
These errors are small in the inner regions of galaxies but can grow to tens of percent as a consequence of the extrapolation to the very large radii of the S\'{e}rsic model, especially when the derived S\'{e}rsic index is large.
If highly precise photometric measurements are required, a description based on the surface brightness distribution within a finite aperture should be preferred over an estimate of the total flux and half-light radius.


\begin{acknowledgements}

I thank Henk Hoekstra for useful discussions.
The Hyper Suprime-Cam (HSC) collaboration includes the astronomical communities of Japan and Taiwan, and Princeton University. The HSC instrumentation and software were developed by the National Astronomical Observatory of Japan (NAOJ), the Kavli Institute for the Physics and Mathematics of the Universe (Kavli IPMU), the University of Tokyo, the High Energy Accelerator Research Organization (KEK), the Academia Sinica Institute for Astronomy and Astrophysics in Taiwan (ASIAA), and Princeton University. Funding was contributed by the FIRST program from the Japanese Cabinet Office, the Ministry of Education, Culture, Sports, Science and Technology (MEXT), the Japan Society for the Promotion of Science (JSPS), Japan Science and Technology Agency (JST), the Toray Science Foundation, NAOJ, Kavli IPMU, KEK, ASIAA, and Princeton University. 

This paper makes use of software developed for the Large Synoptic Survey Telescope. We thank the LSST Project for making their code available as free software at  http://dm.lsst.org

This paper is based [in part] on data collected at the Subaru Telescope and retrieved from the HSC data archive system, which is operated by the Subaru Telescope and Astronomy Data Center (ADC) at National Astronomical Observatory of Japan. Data analysis was in part carried out with the cooperation of Center for Computational Astrophysics (CfCA), National Astronomical Observatory of Japan. The Subaru Telescope is honored and grateful for the opportunity of observing the Universe from Maunakea, which has the cultural, historical and natural significance in Hawaii. 

Funding for the SDSS and SDSS-II has been provided by the Alfred P. Sloan Foundation, the Participating Institutions, the National Science Foundation, the U.S. Department of Energy, the National Aeronautics and Space Administration, the Japanese Monbukagakusho, the Max Planck Society, and the Higher Education Funding Council for England. The SDSS Web Site is http://www.sdss.org/.

The SDSS is managed by the Astrophysical Research Consortium for the Participating Institutions. The Participating Institutions are the American Museum of Natural History, Astrophysical Institute Potsdam, University of Basel, University of Cambridge, Case Western Reserve University, University of Chicago, Drexel University, Fermilab, the Institute for Advanced Study, the Japan Participation Group, Johns Hopkins University, the Joint Institute for Nuclear Astrophysics, the Kavli Institute for Particle Astrophysics and Cosmology, the Korean Scientist Group, the Chinese Academy of Sciences (LAMOST), Los Alamos National Laboratory, the Max-Planck-Institute for Astronomy (MPIA), the Max-Planck-Institute for Astrophysics (MPA), New Mexico State University, Ohio State University, University of Pittsburgh, University of Portsmouth, Princeton University, the United States Naval Observatory, and the University of Washington.

\end{acknowledgements}


\bibliographystyle{aa}
\bibliography{references}

\end{document}